\documentclass[12pt]{article}
\usepackage{multicol}
\usepackage{amssymb}
\usepackage{amsfonts,amssymb,amsmath,amsthm}
\usepackage{epsfig}
\usepackage{slashed}
\usepackage{multirow}
\usepackage{graphicx}
\usepackage{subfig}
\usepackage{mathrsfs}
\usepackage{bbm}
\usepackage{cite}
\usepackage{enumerate}
\usepackage{slashbox}
\usepackage{appendix}
\usepackage{color}
\usepackage[colorlinks,linkcolor=red,anchorcolor=red,citecolor=green]{hyperref}
\makeatletter 

\makeatletter
\@addtoreset{equation}{section}

\newcommand{\Rmnum}[1]{\expandafter\@slowromancap\romannumeral #1@}
\makeatother

\begin{document}
\renewcommand{\thefootnote}{\fnsymbol{footnote}}
\begin{titlepage}

\vspace{10mm}
\begin{center}
{\Large\bf Parametric phase transition for Gauss-Bonnet AdS black hole}
\vspace{10mm}

{{\large Yan-Gang Miao${}^{}$\footnote{\em E-mail: miaoyg@nankai.edu.cn}
and Zhen-Ming Xu}${}^{}$\footnote{\em E-mail: xuzhenm@mail.nankai.edu.cn}

\vspace{3mm}
${}^{}${\normalsize \em School of Physics, Nankai University, Tianjin 300071, China}
}
\end{center}

\vspace{5mm}
\centerline{{\bf{Abstract}}}
\vspace{6mm}
With the help of the parametric solution of the Maxwell equal area law for the Gauss-Bonnet AdS black hole in five dimensions, we find the second analytical solution to the first order phase transition. We analyze the asymptotic behaviors of some characteristic thermodynamic properties for the small and large black holes at the critical and zero temperatures and also calculate the critical exponents and the corresponding critical amplitudes in detail. Moreover, we give the general form of the thermodynamic scalar curvature based on the Ruppeiner geometry and point out that the attractive interaction dominates in both the small and large black hole phases when the first order phase transition occurs in the five dimensional Gauss-Bonnet AdS black hole.

\vspace{5mm}
\noindent
{\bf PACS Number(s)}: 04.70.Dy, 04.70.-s, 05.70.Jk

\vspace{5mm}

\end{titlepage}

\newpage
\renewcommand{\thefootnote}{\arabic{footnote}}
\setcounter{footnote}{0}
\setcounter{page}{2}
\pagenumbering{arabic}
\tableofcontents
\vspace{1cm}

\section{Motivation}
The black hole thermodynamics, which stems largely from the pioneering works by Hawking and Bekenstein\cite{SH,JMB,JDB}, has been increasingly considered as the most feasible topic of semi-classical quantum gravity effect. Especially in the paradigm of extended phase space,\footnote{Here it means that the cosmological constant and its conjugate are treated as the thermodynamic pressure and volume variables, respectively.} the black holes in the AdS spacetime can be made an analogy with the van der Waals fluid and such an analogy can show rich thermodynamic critical phenomena\cite{SHP,RW,TP,SC,CEJM,DSJT,KM,KM1,BPD}. By using the ordinary thermodynamics, one can handle\cite{ESAS,ES1,BC,WL} mainly the first order phase transition of black holes. Unfortunately, it is usually difficult to find an exact and analytical solution to the first order phase transition. Hence, the parametric solution provides\cite{Lekner,LLML} a new possibility, which not only presents the independent variable governing the first order phase transition, but also shows more detailed information of the phase transition. Our current work aims to study the critical behaviors of the Gauss-Bonnet AdS black hole in five dimensions from the viewpoint of parametric solutions.

We summarize the basic formulas of the Gauss-Bonnet AdS black hole for our later use. The action of Einstein-Gauss-Bonnet gravity theory with a negative cosmological constant $\Lambda$ in $d$ dimensions is\cite{RCLY,Zhao}
\begin{align}
{\cal I}=\frac1{16\pi}\int \mathrm{d}^d x \sqrt{-g}\left[R-2\Lambda+\alpha_{_{\mathrm{GB}}} (R_{\mu\nu\gamma\delta}R^{\mu\nu\gamma\delta}-4R_{\mu\nu}R^{\mu\nu}+R^2)\right],
\end{align}
where $\alpha_{_{\mathrm{GB}}}$ is the Gauss-Bonnet coefficient, or known as the second order Lovelock coefficient.\footnote{The Lovelock coefficient is proportional to the inverse string tension in string theory\cite{Deser}. Hence, we take $\alpha_{_{\mathrm{GB}}}>0$ in this paper for the Gauss-Bonnet AdS black hole.}
Correspondingly, the spherically symmetric static solution of the black hole takes\cite{Olea,Astefanesei,Hendi2010,Hendi2016,Wiltshire,RGC,Odintsov,Liu2013,Wang2014,Lan2018,Belhaj2015,Xu2017,Konoplya1,Konoplya2} the form,
\begin{align}
\mathrm{d}s^2=-f(r)\mathrm{d}t^2+\frac{1}{f(r)}\mathrm{d}r^2+r^2 \mathrm{d}\Omega^2,\label{eq:2a}
\end{align}
and
\begin{align}
f(r)=1+\frac{r^2}{2\alpha_0}\left(1-\sqrt{1+\frac{64\pi\alpha_0 {\cal M}}{(d-2)r^{d-1}\Sigma}-\frac{64\pi\alpha_0 P}{(d-1)(d-2)}}\right),
\end{align}
where $\mathrm{d}\Omega^2$ is the square of line element on a $(d-2)$-dimensional maximally symmetric Einstein manifold with volume $\Sigma$. The black hole mass is $\cal M$ and the pressure is $P=-\Lambda/(8\pi)=(d-1)(d-2)/(16\pi l^2)$ with the effective AdS curvature radius $l$. In addition, one can use an auxiliary symbol $\alpha_0\equiv (d-3)(d-4)\alpha_{_{\mathrm{GB}}}$ in order to avoid the verboseness.

In the present paper we will investigate the phase transition and critical behaviors for the Gauss-Bonnet AdS black hole in $d=5$ dimensions. Our main motivations are listed as follows.
\begin{itemize}
  \item Hitherto, a complete theory of quantum gravity has not been established yet. Whatever the quantum gravity theory would be, higher order corrections to the usual Einstein-Hilbert action should exist. The Gauss-Bonnet theory of higher order curvature corrections is an expected candidate. On the other hand, the Gauss-Bonnet theory is a special case of the Lovelock gravity theory. So it is a general second order covariant gravity theory in dimensions higher than four\cite{Lovelock}. According to the explanation of the Gauss-Bonnet coefficient (see footnote 2), the Gauss-Bonnet term can be regarded as corrections from the heterotic string theory\cite{Zumino}. Moreover, the Gauss-Bonnet term represents, from the holographical viewpoint, part of the $1/N$ correction to the large $N$ limit of the dual $SU(N)$-like gauge field theory\cite{Aharony}. Therefore, the investigation of the Gauss-Bonnet AdS black hole is interesting and necessary in its own right.
  \item For the Gauss-Bonnet AdS black hole, the spacetime dimension is $d\geq 5$.\footnote{The Gauss-Bonnet term is just a topological term and the Gauss-Bonnet AdS black hole degenerates into a normal AdS black hole in the case of $d=4$.} In addition, based on the discussion about thermodynamic behaviors of the Gauss-Bonnet AdS black hole\cite{RCLY,Zhao,Xu2017}, one can observe that its thermodynamic behavior in $d=5$ is different from that in other dimensions. Hence, the dimension $d=5$ is of special importance.
  \item Based on the thermodynamics of black holes in the extended phase space, black holes undergo one first order phase transition which is analogous to that of van der Waals fluid\cite{KM}. The Maxwell equal area law is a very useful tool to deal with the first order phase transition\cite{ESAS,WL}. Unfortunately, it is very difficult to obtain the exact analytical solution of the Maxwell equal area law. For black holes, the only known exact analytical solution\cite{ESAS} at present is about RN-AdS black hole in $d=4$ dimensions.\footnote{For the Schwarzschild AdS black hole in $d=4$ dimensions, one can also obtain the exact analytical solution of the Maxwell equal area law\cite{ES1,BC}, where the function of the Maxwell equal area law is used to remove the negative heat capacity rather than to analyze the first order phase transition.} In addition, the corresponding parametric analytical solution is also known\cite{LLML}. For other types of black holes, the numerical methods are often used to analyze phase transitions. Fortunately, we work out the parametric analytical solution of the Maxwell equal area law for the Gauss-Bonnet AdS black hole in $d=5$ dimensions, which may be thought of as the second analytical solution.
  \item With the help of the parametric analytical solution to the first order phase transition, we can analyze in detail the critical behaviors at the critical and zero temperatures for the Gauss-Bonnet AdS black hole in $d=5$ dimensions. At the same time, we can also calculate directly the critical exponents and the corresponding critical amplitudes.
\end{itemize}

The paper is organized as follows. In section \ref{sec2}, we briefly review some basic thermodynamic properties and the formula of the Maxwell equal area for the Gauss-Bonnet AdS black hole in $d=5$ dimensions. In section \ref{sec3}, we give the parametric analytical solution of the Maxwell equal area law for the Gauss-Bonnet AdS black hole in $d=5$ dimensions. We then use the parametric solution to analyze in detail the critical behaviors at the critical and zero temperatures in section \ref{sec4}. Finally, we devote to drawing our conclusion in section \ref{sec5}. In addition, the general form of the thermodynamic scalar curvature based on the Ruppeiner geometry is  attached in Appendix \ref{app}. This scalar curvature is calculated for the study of interaction in the small and large black hole phases from the viewpoint of Ruppeiner thermodynamic geometry. Throughout this paper, we adopt the units $\hbar=c=k_{_{B}}=G=1$.

\section{Thermodynamics for Gauss-Bonnet AdS black hole in $d=5$ dimensions}\label{sec2}
The basic thermodynamic properties of the Gauss-Bonnet AdS  black hole take the following forms in terms of the horizon radius $r_h$ which is determined by the zero point of $g^{-1}_{rr}$ component in the metric\cite{RGC,Odintsov,Liu2013,Wang2014,Lan2018,Belhaj2015,Xu2017},
\begin{eqnarray}
\text{Enthalpy}&:&M=\frac{3\pi r_h^2}{8}\left(1+\frac{\alpha_0}{r_h^2}+\frac{4\pi P r_h^2}{3}\right),\\
\text{Temperature}&:&T=\frac{8\pi P r_h^3+3r_h}{6\pi(r_h^2+2\alpha_0)},\label{temp}\\
\text{Entropy}&:&S=\frac{\pi^2 r_h^3}{2}\left(1+\frac{6\alpha_0}{r_h^2}\right),\label{entr}\\
\text{(Thermo)Volume}&:&V=\frac{\pi^2 r_h^4}{2},\label{volu}\\
\text{Equation of state}&:&P(r_h,T)=\frac{3T}{4r_h}\left(1+\frac{2\alpha_0}{r_h^2}\right)-\frac{3}{8\pi r_h^2}.\label{eos}
\end{eqnarray}

\begin{figure}
\begin{center}
\includegraphics[width=85mm]{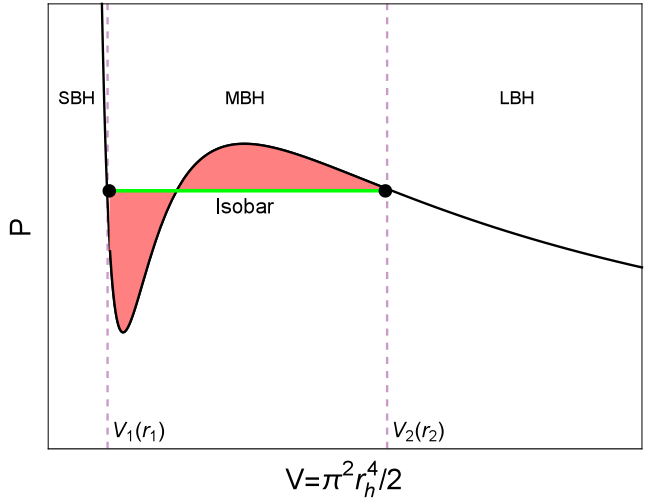}
\end{center}
\caption{At $\alpha_0=1$ and $T=0.058 <T_c$, the schematic diagram of the $P-V$ (or $P-r_h$) graph described by eq.~(\ref{eos}). The abbreviations LBH, SBH and MBH refer to large, small, and medium-sized black holes, respectively. $V_1$ and $V_2$ denote the (thermo)volume for the small and large black holes with the horizon radii $r_1$ and $r_2$, respectively. The Maxwell equal area law requires that the areas of the two shaded parts are equal to each other.}
\label{fig0}
\end{figure}

At a constant temperature with fixed $\alpha_0$, the schematic diagram of the equation of state (eq.~(\ref{eos})), i.e. the pressure with respect to the (thermo)volume, the $P-V$ graph is shown in Figure~\ref{fig0}. 

We can clearly see that there exist two inflection points in the $P-V$ graph, i.e. the local minimum and local maximum points. When the two inflection points merge into one, this phenomenon is often referred to as the $P-V$ criticality. Correspondingly, the critical values are determined by the following relations,
\begin{equation}
\frac{\partial P}{\partial r_h}=0, \qquad \frac{\partial^2 P}{\partial r_h^2}=0. \label{inf}
\end{equation}
For the Gauss-Bonnet AdS black hole in $d=5$ dimensions, one can compute~\cite{RGC,Odintsov,Liu2013,Wang2014,Lan2018,Belhaj2015,Xu2017} these critical values by using eqs.~(\ref{eos}) and~(\ref{inf}), 
\begin{equation}
r_c=\sqrt{6\alpha_0}, \quad V_c=18\pi^2 \alpha_0^2, \quad T_c=\frac{1}{2\pi\sqrt{6\alpha_0}}, \quad P_c=\frac{1}{48\pi \alpha_0}, \quad S_c=6\pi^2 \alpha_0\sqrt{6\alpha_0}. \label{cv}
\end{equation}

Below these critical values, there is an oscillating part in the $P-V$ graph (see Figure \ref{fig0}). In other words, there exists three black hole phases with different sizes at the same pressure when $T<T_c$. The three sizes of black holes are marked by large, small, and medium-sized black holes, respectively. The medium-sized black hole corresponds to an oscillating part in the $P-V$ graph and is unstable due to its negative heat capacity at constant pressure. Hence, we have to utilize the Maxwell equal area law to replace this oscillating part by an isobar, which implies that the small black hole can jump to the large one through the isobaric process. This phenomenon is called the first order phase transition. The oscillating part, i.e. the medium-sized black hole, corresponds to the curve between two black points in Figure \ref{fig0}. The Maxwell equal area law requires that the areas above and below the isobar are equal to each other. As a result, one can obtain~\cite{Xu2017} the following equations,
\begin{equation}
\begin{aligned}
& P(r_1,T)=P=P(r_2,T),\\
& P\cdot(V_2-V_1)=\int_{r_1}^{r_2}P(r_h,T)\mathrm{d}V, \label{meal}
\end{aligned}
\end{equation}
where $P$ stands for an isobar, and $V_1$ and $V_2$ denote the (thermo)volume defined by eq.~(\ref{volu}) for the small and large black holes with the horizon radii $r_1$ and $r_2$, respectively. In the next section we will parameterize the solution of the equation (\ref{meal}) and analyze its relevant critical behaviors.

\section{Parametric phase transition}\label{sec3}
Substituting eq.~(\ref{eos}) into eq.~(\ref{meal}), 
we obtain a key expression\footnote{This expression first appeared in ref.\cite{Xu2017}, where it was used to numerically analyze the phase transition and latent heat.} for parametric phase transition,
\begin{eqnarray}
r_1 r_2=6\alpha_0. \label{cent}
\end{eqnarray}
It is the result that allows us to parameterize the phase transition for the Gauss-Bonnet AdS black hole. The specific process is as follows.

Based on eq.~(\ref{cent}), it is crucial to introduce a dimensionless auxiliary parameter $y$ that parameterize\cite{Xu2017} the horizon radii of the small and large black holes in the following way,
\begin{eqnarray}
r_1=y\sqrt{6\alpha_0}, \qquad r_2=\frac{\sqrt{6\alpha_0}}{y}. \label{pcent}
\end{eqnarray}
Some dimensionless reduced parameters are usually introduced for the sake of convenience,
\begin{eqnarray}
\hat{t}:=\frac{T}{T_c}, \qquad \hat{s}:=\frac{S}{S_c}, \qquad \hat{p}:=\frac{P}{P_c}, \qquad \hat{v}:=\frac{V}{V_c}, \qquad \hat{n}:=\frac{1}{\hat{v}},\label{redp}
\end{eqnarray}
where $\hat{n}$ is the dimensionless reduced number density,\footnote{Here $N$ is the total number of molecules in the black hole system, and the number density is defined by $n\equiv N/V$. In addition, the critical number density is $n_c=N/V_c$, so we obtain the dimensionless reduced number density $\hat{n}:=n/n_c=1/\hat{v}$.} and $0\leq \hat{t}\leq 1$ and $0\leq \hat{p}\leq 1$. Inserting eq.~(\ref{pcent}) into eq.~(\ref{meal}) and considering eq.~(\ref{redp}), we derive the dimensionless reduced temperature and pressure of parametric first order phase transition for the Gauss-Bonnet AdS black hole,
\begin{eqnarray}
\hat{t}=\frac{3(y^3+y)}{y^4+4y^2+1}, \qquad \hat{p}=\frac{6y^2}{y^4+4y^2+1}.\label{tp}
\end{eqnarray}
In this way, we can give a relationship between the thermodynamic situations and the limits of the auxiliary parameter $y$. Because of $r_2\geq r_1$, we see $0\leq y\leq 1$. More precisely, we can obtain
\begin{align}
\text{Critical situation:}& \qquad \hat{t}\rightarrow 1 \qquad \Leftrightarrow \qquad y\rightarrow 1 \label{crcase}\\
\text{Extreme situation:}& \qquad \hat{t}\rightarrow 0 \qquad \Leftrightarrow \qquad y\rightarrow 0 \label{excase}
\end{align}
For the small black hole phase, we have
\begin{eqnarray}
\hat{v}_1=y^4, \qquad \hat{s}_1=\frac12(y^3+y), \qquad \hat{n}_1=\frac{1}{y^4}, \label{smallphase}
\end{eqnarray}
and for the large black hole phase, we have
\begin{eqnarray}
\hat{v}_2=\frac{1}{y^4}, \qquad \hat{s}_2=\frac12\left(\frac{1}{y^3}+\frac{1}{y}\right), \qquad \hat{n}_2=y^4. \label{largephase}
\end{eqnarray}
At the same time, due to eq.~(\ref{tp}) we can also get the slope of the co-existence curve of the small and large black hole phases,
\begin{eqnarray}
\frac{\mathrm{d}\hat{p}}{\mathrm{d}\hat{t}}=\frac{4(y^3+y)}{y^4+1}.
\end{eqnarray}
Furthermore, when the first order phase transition occurs, i.e. the small black hole phase jumps to the large one, we can give the difference in volume, entropy and number density between the small and large black hole phases at the two edges of the co-existence curve,
\begin{subequations}
\begin{align}
\Delta\hat{v} &:=\hat{v}_2-\hat{v}_1=\frac{1}{y^4}-y^4,\\
\Delta\hat{s} &:=\hat{s}_2-\hat{s}_1=\frac12\left(\frac{1}{y^3}+\frac{1}{y}-y^3-y\right),\\
\Delta\hat{n} &:=\hat{n}_1-\hat{n}_2=\frac{1}{y^4}-y^4,
\end{align}
\end{subequations}
and the latent heat $L$ of the first order phase transition on crossing the small-large black hole phase co-existence curve,
\begin{equation}
\frac{L}{P_c V_c}=\frac{12(1+y^2)^2(1-y^4)}{y^2(y^4+4y^2+1)}, \label{late}
\end{equation}
where the Clausius-Clapeyron equation has been used.\footnote{Considering the Clausius-Clapeyron equation\cite{DCJ},
\begin{equation*}
\frac{\mathrm{d}P}{\mathrm{d}T}=\frac{L}{T \Delta V} \qquad \text{or} \qquad \frac{\mathrm{d}P}{\mathrm{d}T}=\frac{\Delta S}{\Delta V},
\end{equation*}
and the dimensionless reduced parameters in eq.~(\ref{redp}), we can write the latent heat,
\begin{equation*}
\frac{L}{P_c V_c}=\hat{t}\Delta\hat{v}\frac{\mathrm{d}\hat{p}}{\mathrm{d}\hat{t}} \qquad \text{or}\qquad  \frac{L}{T_c S_c}=\hat{t}\Delta\hat{s}.
\end{equation*}
Again using eq.~(\ref{cv}), we have the relation: $\frac{L}{P_c V_c}=\frac{8L}{T_c S_c}$.
}
Finally, we compute the reduced thermodynamic scalar curvatures\footnote{See Appendix \ref{app} for the details.} of the small and large black hole phases for the Gauss-Bonnet AdS black hole in $d=5$ dimensions,
\begin{subequations}
\begin{align}
\hat{\mathcal{R}}_1&:=\frac{\mathcal{R}_1}{|\mathcal{R}_c|}=-\frac{16(y^4+4y^2+1)}{3y(y^2+1)(3y^2+1)^2},\label{smallr}\\
\hat{\mathcal{R}}_1&:=\frac{\mathcal{R}_2}{|\mathcal{R}_c|}=-\frac{16y^3(y^4+4y^2+1)}{3(y^2+1)(y^2+3)^2}. \label{larger}
\end{align}
\end{subequations}

\begin{figure}
\begin{center}
  \begin{tabular}{cc}
    \includegraphics[width=65mm]{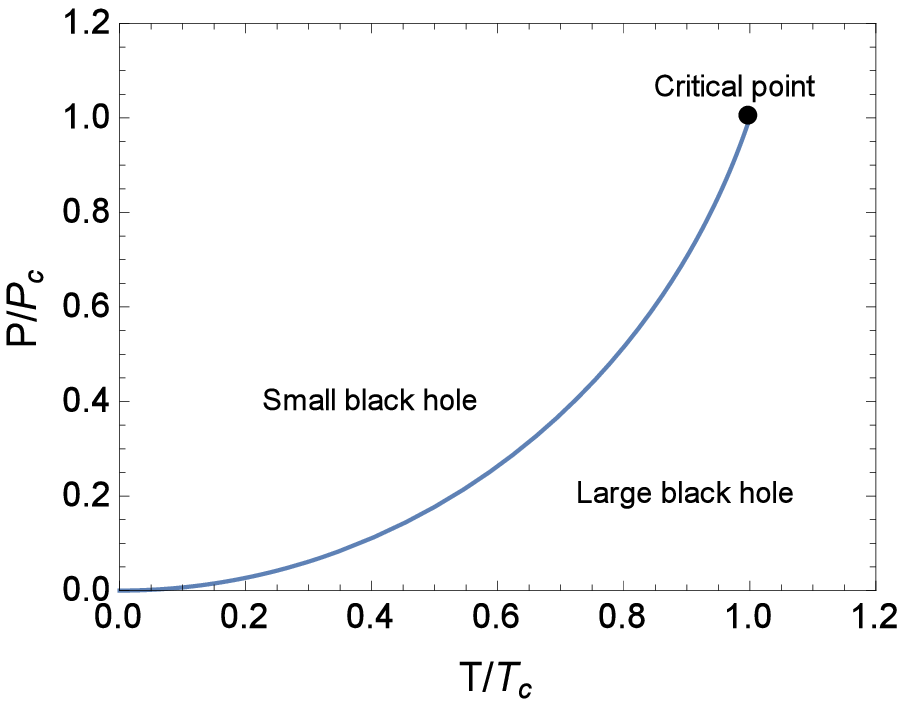} &
    \includegraphics[width=65mm]{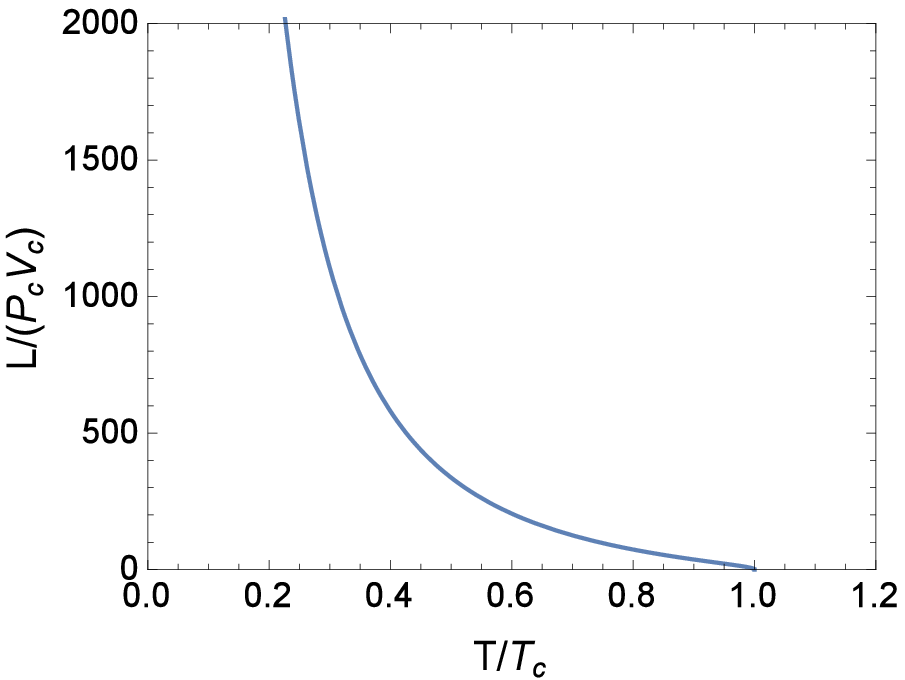} \\
    \includegraphics[width=65mm]{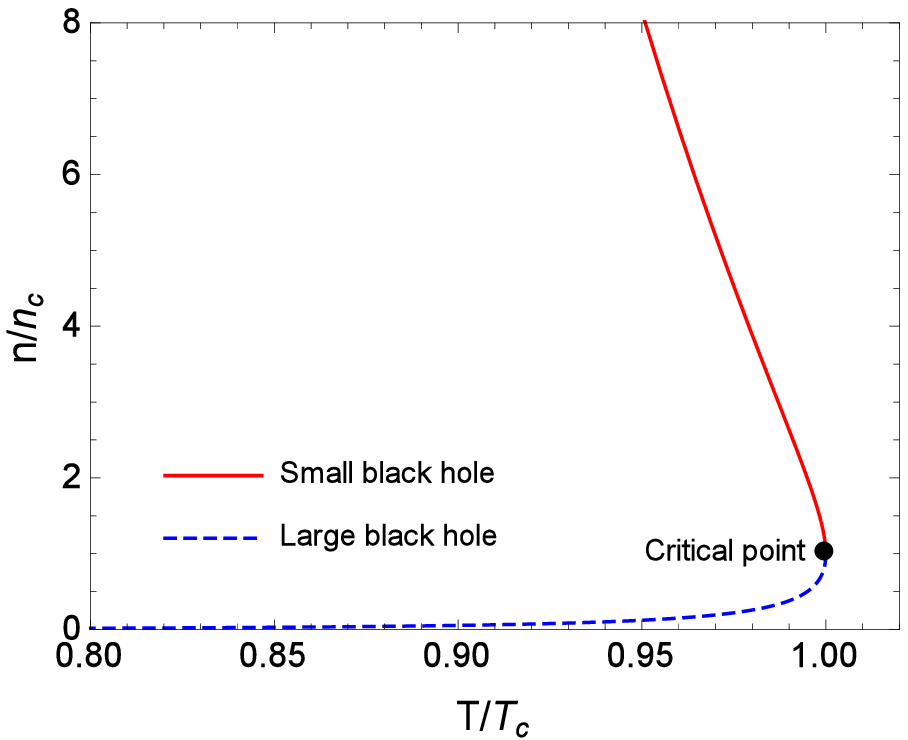} &
    \includegraphics[width=65mm]{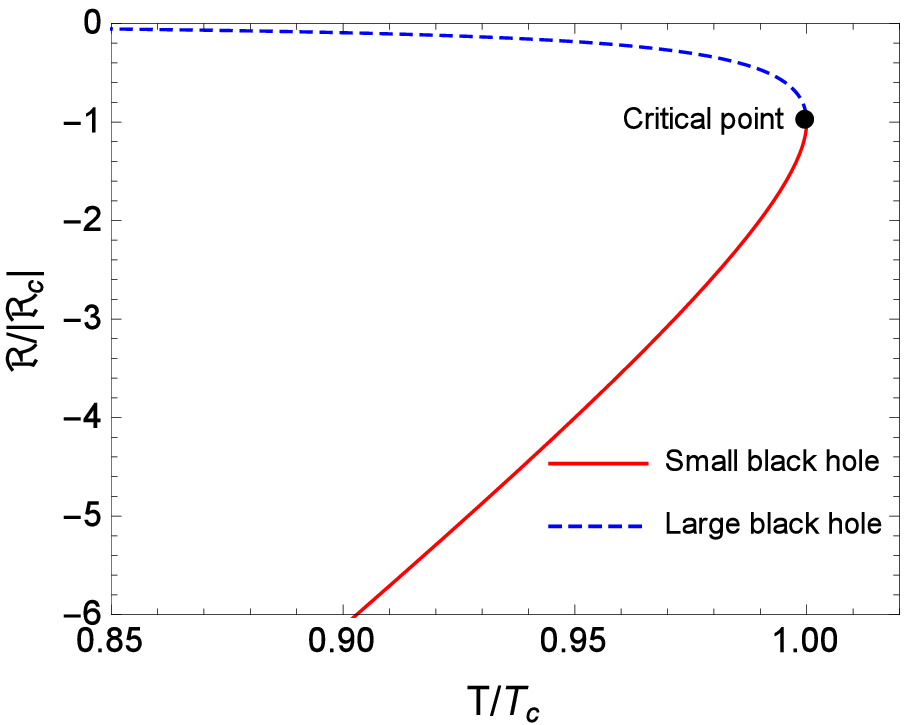}
  \end{tabular}
\end{center}
\caption{The behaviors of some thermodynamic quantities with respect to the temperature for the five dimensional Gauss-Bonnet AdS black hole. {\bf Left top:} the co-existence curve of small and large black hole phases. {\bf Left bottom:} the number densities of the small and large black holes on the co-existence curve. {\bf Right top:} the latent heat on crossing the small-large black hole phase co-existence curve. {\bf Right bottom:} the thermodynamic scalar curvatures of small and large black hole phases.}
\label{fig}
\end{figure}

Figure \ref{fig} shows the co-existence curve of small and large black hole phases (described by eq.~(\ref{tp})), the latent heat on crossing the small-large black hole phase co-existence curve (described by eq.~(\ref{late})), and the number densities and the thermodynamic scalar curvatures of small and large black hole phases with respect to the temperature, respectively. 

The co-existence curve begins at the origin and terminates at the critical point. Above the critical point, we cannot distinguish the two phases of small and large black holes. The latent heat goes to infinity with the temperature tending to zero and it approaches zero with the temperature tending to the critical value. This is consistent with the result that was numerically calculated in ref.\cite{Xu2017}. For the large black hole, its number density increases with the temperature increasing. However, for the small black hole, its number density decreases with the temperature increasing and it tends to infinity when the temperature goes to zero. Both the thermodynamic scalar curvatures of small and large black holes are negative, which implies that the attractive interaction dominates. Meanwhile, we notice that the thermodynamic scalar curvature of the small black hole will go to negative infinity with the temperature tending to zero. As to the case at the critical or zero temperature, the critical behaviors of the small and large black hole phases will be discussed in the next section.

\section{Parametric critical behavior}\label{sec4}
\subsection{Approaching the critical temperature}
In order to investigate more clearly the thermodynamic behaviors of the small and large black hole phases at the critical temperature, we define the following two parameters,
\begin{equation}
\tau \equiv 1-\hat{t}, \qquad x\equiv 1-y, \label{sym}
\end{equation}
where they satisfy the inequalities: $0\leq \tau \leq 1$ and $0\leq x \leq 1$. From eq.~(\ref{crcase}), we see that $\tau\rightarrow 0$ and $x\rightarrow 0$ in the critical situation. Substituting eq.~(\ref{sym}) into the reduced temperature eq.~(\ref{tp}) and making the Taylor expansion at $x=0$, we obtain
\begin{equation}
\tau=\frac{x^2}{6}+\frac{x^3}{6}+\frac{2x^4}{9}+\mathcal{O}(x^5). \label{tau}
\end{equation}
Then we solve $x$ reversely from the above relation,
\begin{equation}
x=\sqrt{6}\tau^{1/2}-3\tau-\sqrt{\frac{3}{8}}\tau^{3/2}+\mathcal{O}(\tau^2). \label{xtau}
\end{equation}

\subsubsection{Small black hole at the critical temperature}
Substituting eq.~(\ref{sym}) into eq.~(\ref{smallphase}) and making the Taylor expansion at $x=0$, and considering eq.~(\ref{xtau}) again, we obtain
\begin{subequations}\label{sbh}
\begin{align}
\hat{v}_1-1 &=-4\sqrt{6}\tau^{1/2}+48\tau-59\sqrt{6}\tau^{3/2}+\mathcal{O}(\tau^2),\label{sbhv}\\
\hat{s}_1-1 &=-2\sqrt{6}\tau^{1/2}+15\tau-23\sqrt{\frac{3}{2}}\tau^{3/2}+\mathcal{O}(\tau^2),\label{sbhs}\\
\hat{n}_1-1 &=4\sqrt{6}\tau^{1/2}+48\tau+59\sqrt{6}\tau^{3/2}+\mathcal{O}(\tau^2), \label{sbhn}
\end{align}
\end{subequations}
and the reduced thermodynamic scalar curvature,
\begin{equation}
\hat{\mathcal{R}}_1+1=-3\sqrt{6}\tau^{1/2}-\frac{47\tau}{2}-\frac{45}{2}\sqrt{\frac{3}{2}}\tau^{3/2}+\mathcal{O}(\tau^2).\label{sbhr}
\end{equation}

\subsubsection{Large black hole at the critical temperature}
Making a similar discussion to that in the above subsection except for substituting eq.~(\ref{sym}) into eq.~(\ref{largephase}), we eventually have
\begin{subequations}\label{lbh}
\begin{align}
\hat{v}_2-1 &=4\sqrt{6}\tau^{1/2}+48\tau+59\sqrt{6}\tau^{3/2}+\mathcal{O}(\tau^2),\label{lbhv}\\
\hat{s}_2-1 &=2\sqrt{6}\tau^{1/2}+15\tau+23\sqrt{\frac{3}{2}}\tau^{3/2}+\mathcal{O}(\tau^2),\label{lbhs}\\
\hat{n}_2-1 &=-4\sqrt{6}\tau^{1/2}+48\tau-59\sqrt{6}\tau^{3/2}+\mathcal{O}(\tau^2), \label{lbhn}
\end{align}
\end{subequations}
and the reduced thermodynamic scalar curvature,
\begin{equation}
\hat{\mathcal{R}}_2+1=3\sqrt{6}\tau^{1/2}-\frac{47\tau}{2}+\frac{45}{2}\sqrt{\frac{3}{2}}\tau^{3/2}+\mathcal{O}(\tau^2).\label{lbhr}
\end{equation}

Here we make two comments to eqs.~(\ref{sbh}), (\ref{sbhr}), (\ref{lbh}) and (\ref{lbhr}). The first is that the thermodynamic behaviors of the small and large black hole phases are symmetrically distributed on both sides of the critical point. The other comment is that the leading order term related to $\tau^{1/2}$ in each of the equations will play a major role in calculating the static critical exponents  in the next subsection.

\subsubsection{Static critical exponent}
Critical exponents describe the behavior of thermodynamic quantities when the temperature approaching phase transition points
 and they do not depend on the details of physical systems, i.e., they are universal\cite{KM,DCJ}. For the sake of calculation of critical exponents, one usually introduces the following notations,
\begin{equation}
\tau_0:=-\tau:=\hat{t}-1, \qquad v_0:=\hat{v}-1, \qquad n_0:=\hat{n}-1, \qquad p_0:=\hat{p}-1. \label{nota}
\end{equation}
\begin{itemize}
  \item $\alpha$ and $\alpha^{\prime}$ are related to the specific heat at constant volume:
  \begin{equation}
  \begin{split}
  C_V&=a\tau_0^{-\alpha}, \quad \,~\text{when} \quad \hat{t}\rightarrow 1^{+},   \\
  C_V&=a^{\prime}\tau^{-\alpha^{\prime}}, \quad \text{when} \quad \hat{t}\rightarrow 1^{-},
  \end{split}
  \end{equation}
  where $a$ and $a^{\prime}$ are the critical amplitudes that correspond to the critical exponents $\alpha$ and $\alpha^{\prime}$, respecttively. For the Gauss-Bonnet AdS black hole, due to $C_V=0$, it is easy to have
  \begin{equation}
  \alpha=\alpha^{{\prime}}=0, \qquad a=a^{{\prime}}=0.
  \end{equation}

  \item $\beta$ is related to the difference between the small and large black hole number densities on crossing the co-existence curve:
  \begin{equation}
  \Delta\hat{n}=\hat{n}_1-\hat{n}_2=b \tau^{\beta}, \quad \text{when} \quad \hat{t}\rightarrow 1^{-}.
  \end{equation}
 Note that one cannot define a critical exponent and its amplitude for the path $\hat{t}\rightarrow 1^{+}$ since the order parameter $\Delta\hat{n}$ is zero for $\hat{t}>1$. According to eqs.~(\ref{sbhn}) and ~(\ref{lbhn}), we obtain $\Delta\hat{n}=8\sqrt{6}\tau^{1/2}$. So the critical exponent and amplitude are
  \begin{equation}
  \beta=\frac12, \qquad b=8\sqrt{6}.
  \end{equation}

  \item $\gamma$ is related to the isothermal compressibility:
  \begin{equation}
  \begin{split}
  \kappa_{_T}P_c&=g\tau_0^{-\gamma}, \quad \,~\text{when} \quad \hat{t}\rightarrow 1^{+},   \\
  \kappa_{_T}P_c&=g^{{\prime}}\tau^{-\gamma^{{\prime}}}, \quad \text{when} \quad \hat{t}\rightarrow 1^{-},
  \end{split}
  \end{equation}
  where $\kappa_{_T}$ is isothermal compression coefficient. By means of inserting eqs.~(\ref{redp}) and~(\ref{nota}) into eqs.~(\ref{volu}) and~(\ref{eos}), we get the following critical behavior,
  \begin{equation}
  \frac{1}{\kappa_{_T}P_c}=-\left.\hat{v}\left(\frac{\partial \hat{p}}{\partial\hat{v}}\right)_{\hat{t}}\right|_{\hat{v}\rightarrow 1, \hat{t}\rightarrow 1^{+}}=\frac32 \tau_0.
  \end{equation}
  Thus we directly read the critical exponent and its amplitude,
  \begin{equation}
  \gamma=1, \qquad g=\frac{2}{3}.
  \end{equation}
  In addition, for the path $\hat{t}\rightarrow 1^{-}$, substituting eqs.~(\ref{sym}) and~(\ref{nota}) into the reduced pressure eq.~(\ref{tp}) and making the Taylor expansion at $x=0$, we obtain $p_0=-4\tau$ with the help of eq.~(\ref{xtau}). Next, considering eqs.~(\ref{sbhv}) and~(\ref{lbhv}), we have $v^2_0=(4\sqrt{6})^2 \tau$. Combing the relations of  $p_0$ and  $v^2_0$, we deduce $p_0=-v_0^2/24$. As a result, we derive the desired critical behavior,
  \begin{equation}
  \frac{1}{\kappa_{_T}P_c}=-\left.v_0\left(\frac{\partial p_0}{\partial v_0}\right)_{\hat{t}}\right|_{\hat{t}\rightarrow 1^{-}}=8 \tau,
  \end{equation}
  from which the critical exponent and amplitude can be read
  \begin{equation}
  \gamma^{{\prime}}=1, \qquad g^{{\prime}}=\frac{1}{8}.
  \end{equation}

  \item $\delta$ is related to the critical behavior in the critical isotherm $\hat{t}=1$:
  \begin{equation}
  p_0=d |n_0|^{\delta}, \quad \text{when} \quad \hat{n}\rightarrow 1^{+,-}.
  \end{equation}
  Inserting eqs.~(\ref{redp}) and~(\ref{nota}) into eqs.~(\ref{volu}) and~(\ref{eos}), we have $p_0=-n_0^3/64$. Hence, the critical exponent and amplitude are
  \begin{equation}
  \delta=3, \qquad d=\frac{1}{64}.
  \end{equation}
\end{itemize}

Here we list the critical exponents and critical amplitudes about the van der Waals fluid and Gauss-Bonnet AdS black hole in $d=5$ dimensions in Table \ref{tab1}. One can see that the critical exponents are same for the two systems, which implies that the critical exponents are universal. However, the critical amplitudes of the two systems are different from each other, which shows that the critical amplitudes present characteristics of systems, i.e. they can be used to distinguish different thermodynamic systems.

\begin{table}[!htbp]
\centering
\begin{tabular}{|c|c|c|c|c|c|}
\hline
\multirow{2}*{Exponent}&\multirow{2}*{Definition}& \multicolumn{2}{c|}{van der Waals fluid}&\multicolumn{2}{c|}{Gauss-Bonnet AdS BH}\\
\cline{3-6}
\multicolumn{1}{|c|}{}&\multicolumn{1}{|c|}{}&Exponent&Amplitude&Exponent&Amplitude\\
\hline
$\alpha$     &$C_V=a\tau_0^{-\alpha}$  &$\alpha=0$ &$a=\frac32$ &$\alpha=0$ &$a=0$\\
$\alpha^{{\prime}}$ &$C_V=a^{{\prime}}\tau^{-\alpha^{{\prime}}}$ &$\alpha^{{\prime}}=0$ &$a^{{\prime}}=6$ &$\alpha^{{\prime}}=0$ &$a^{{\prime}}=0$\\
$\beta$      &$\Delta\hat{n}=b \tau^{\beta}$ &$\beta=\frac12$ &$b=4$ &$\beta=\frac12$ &$b=8\sqrt{6}$\\
$\gamma$     &$\kappa_{_T}P_c=g\tau_0^{-\gamma}$ &$\gamma=1$ &$g=\frac{1}{6}$ &$\gamma=1$ &$g=\frac{2}{3}$\\
$\gamma^{{\prime}}$ &$\kappa_{_T}P_c=g^{{\prime}}\tau^{-\gamma^{{\prime}}}$ &$\gamma^{{\prime}}=1$ &$g^{{\prime}}=\frac{1}{12}$ &$\gamma^{{\prime}}=1$ &$g^{{\prime}}=\frac{1}{8}$\\
$\delta$     &$p_0=d |n_0|^{\delta}$ &$\delta=3$ &$d=\frac32$ &$\delta=3$ &$d=\frac{1}{64}$\\
\hline
\end{tabular}
\caption{Comparison of the critical exponents and critical amplitudes between the van der Waals fluid and the Gauss-Bonnet AdS black hole in $d=5$ dimensions. The data of the van der Waals fluid stem from ref.\cite{DCJ}.}
\label{tab1}
\end{table}

\subsection{Approaching zero temperature}
The thermodynamic behaviors of the small and large black hole phases at zero temperature for the Gauss-Bonnet AdS black hole in $d=5$ dimensions will be discussed in detail in this subsection. From eq.~(\ref{excase}), we see that for studying the critical behaviors at zero temperature we need to make the Taylor expansion for the reduced temperature eq.~(\ref{tp}) at $y=0$,
\begin{equation}
\hat{t}=3y-9y^3+33y^5+\mathcal{O}(y^7), \label{zerot}
\end{equation}
and then we solve $y$ reversely from the above relation,
\begin{equation}
y=\frac{\hat{t}}{3}+\frac{\hat{t}^3}{9}+\frac{16\hat{t}^5}{243}+\mathcal{O}(\hat{t}^7). \label{zeroy}
\end{equation}
Next, when inserting eq.~(\ref{zeroy}) into eqs.~(\ref{smallphase}) and~(\ref{largephase}), we can obtain the thermodynamic critical characteristics of the small and large black hole phases at zero temperature for the Gauss-Bonnet AdS black hole in $d=5$ dimensions. We list them as follows.

The reduced (thermo)volumes of the small and large black hole phases:
\begin{subequations}
\begin{align}
\hat{v}_1&=\frac{\hat{t}^4}{81}+\frac{4\hat{t}^6}{243}+\frac{118\hat{t}^8}{243}+\mathcal{O}(\hat{t}^{10}),\\
\hat{v}_2&=\frac{81}{\hat{t}^4}+\frac{108}{\hat{t}^2}+26+\mathcal{O}(\hat{t}^2).
\end{align}
\end{subequations}

The reduced entropy:
\begin{subequations}
\begin{align}
\hat{s}_1&=\frac{\hat{t}}{6}+\frac{2\hat{t}^3}{27}+\frac{25\hat{t}^5}{486}+\mathcal{O}(\hat{t}^7),\\
\hat{s}_2&=\frac{27}{2\hat{t}^3}-\frac{12}{\hat{t}}+\frac{\hat{t}}{2}+\mathcal{O}(\hat{t}^3),
\end{align}
\end{subequations}

The reduced number densities:
\begin{subequations}
\begin{align}
\hat{n}_1&=\frac{81}{\hat{t}^4}+\frac{108}{\hat{t}^2}+26+\mathcal{O}(\hat{t}^2),\\
\hat{n}_2&=\frac{\hat{t}^4}{81}+\frac{4\hat{t}^6}{243}+\frac{118\hat{t}^8}{243}+\mathcal{O}(\hat{t}^{10}).
\end{align}
\end{subequations}

The reduced thermodynamic scalar curvatures:
\begin{subequations}
\begin{align}
\hat{\mathcal{R}}_1&=-\frac{16}{\hat{t}}+\frac{32\hat{t}}{3}+\frac{16\hat{t}^3}{9}+\mathcal{O}(\hat{t}^5),\\
\hat{\mathcal{R}}_2&=-\frac{16\hat{t}^3}{729}-\frac{544\hat{t}^5}{19683}+\mathcal{O}(\hat{t}^7).
\end{align}
\end{subequations}

From the above formulas, we can clearly see that when the temperature approaches zero, the critical values of the number density and thermodynamic scalar curvature indeed go to infinity for the small black hole, but the two values go to zero for the large black hole. The reason is that the (thermo)volume or entropy of small (large) black hole goes to zero (infinity) when the temperature approaches zero. The analytic analysis is consistent with the description of Figure \ref{fig}.

\section{Summary}\label{sec5}
Based on the parametric analytical solution to the first order phase transition for the Gauss-Bonnet AdS black hole in $d=5$ dimensions, we investigate the thermodynamic critical behaviors. This parametric solution may be regarded as the second analytic solution of the Maxwell equal area law. Some comments are given on similarities and differences between our parametric analytical solution and the only solution so far existing in the literature~\cite{LLML}.
\begin{itemize}
  \item Both results obtained in the present paper and in ref.~\cite{LLML} are based on the Maxwell equal area law. Using such a parametric analytical solution to the first order phase transition, one can get more information about the thermodynamic critical behaviors of black holes, such as the critical exponents, critical amplitudes and interaction.
  \item The difference of entropy between the small and large black hole phases was taken~\cite{LLML} as an auxiliary parameter in order to achieve the parametric analytical solution. Nonetheless, the dimensionless auxiliary parameter $y$ related to the Gauss-Bonnet coefficient is introduced in the present paper in order to realize the parametric solution.
  \item The RN-AdS black hole in $d=4$ dimensions was dealt with in ref.~\cite{LLML}. For this model, its exact analytical (non-parametric) solution of the Maxwell equal area law had been known in ref.~\cite{ESAS}. For the Gauss-Bonnet AdS black hole, however, it is very difficult to obtain the exact analytical solution in arbitrary dimensions. Fortunately, we find one parametric analytical solution in $d=5$ dimensions, which might provide a new perspective for further studies of thermodynamic behaviors of the Gauss-Bonnet AdS black hole.
\end{itemize}

Furthermore, at the critical temperature, we have calculated the critical exponents and the corresponding critical amplitudes in detail. By comparing them with that of the van der Waals fluid, we can see clearly that the critical exponents are universal, but the critical amplitudes present characteristics of thermodynamic systems. By means of the thermodynamic scalar curvature, we have obtained that the attractive interaction dominates in both the small and large black hole phases. In addition, we have also analyzed the asymptotic behaviors of thermodynamic properties for the small and large black holes at zero temperature.
With the help of the parametric solution of the Maxwell equal area law, we have acquired more details about the first order phase transition of black holes. Our treatment can be extended to other black hole models and the key point is to select such an appropriate auxiliary parameter that an analytical solution can be found.

\section*{Acknowledgments}
This work was supported in part by the National Natural Science Foundation of China under grant No.11675081. The authors would like to thank the anonymous referees for the helpful comments that improve this work greatly.

\appendix
\section{Thermodynamic scalar curvature}\label{app}
The Ruppeiner geometry is a powerful tool to investigate microscopic properties of black holes completely from the thermodynamic point of view\cite{GR0,GR1,GR2,GR3,GR4,GR5,GR6,GR7,Mansoori2014,Mansoori2015,Mansoori2016,Hendi2015,GR8,GR9,Vetsov}. Its metric can be written in the Weinhold energy form~\cite{FW},
\begin{equation}
g_{\alpha\beta}=\frac{1}{k_{_B}T_h}\frac{\partial^2 M}{\partial X^{\alpha}\partial X^{\beta}}, \label{rg}
\end{equation}
where $X^{\alpha}=(S,P)$. The first law of black hole thermodynamics takes the following form,
\begin{equation}
\mathrm{d}M=T\mathrm{d}S+V\mathrm{d}P+\cdots. \label{flaw}
\end{equation}
Note that both the enthalpy $M$ and temperature $T$ are linearly dependent on pressure $P$, which implies $\partial^2 M/\partial P^2=0$ and $\partial^2 T/\partial P^2=0$. Based on the facts, one can obtain the general form of the thermodynamic scalar curvature induced by eq.~(\ref{rg}),
\begin{equation}
\mathcal{R}=\frac{\partial}{\partial S}\left[\ln\left(\left. T \middle/\frac{\partial T}{\partial P} \right.\right)\right]. \label{tcurvature}
\end{equation}

Importantly, the thermodynamic scalar curvature can qualitatively reflect some information of the internal interaction of a thermodynamic system. That is, a positive (negative) thermodynamic scalar curvature implies a repulsive (attractive) interaction, while a vanishing thermodynamic scalar curvature implies no interaction\cite{GR2,GR3}.

Naturally, we compute the thermodynamic scalar curvature for the Gauss-Bonnet AdS black hole in $d=5$ dimensions with the help of eqs.~(\ref{temp}) and~(\ref{entr}),
\begin{equation}
\mathcal{R}=-\frac{4}{\pi^2 r_h(r_h^2+2\alpha_0)(8\pi P r_h^2+3)}. \label{gbcur}
\end{equation}
At the critical point eq.~(\ref{cv}), the critical thermodynamic scalar curvature takes the form,
\begin{equation}
\mathcal{R}_c=-1/(8\pi^2 \alpha_0\sqrt{6\alpha_0}).
\end{equation}

\end{document}